\documentclass{article}
  \usepackage{epsfig}
    
\begin{document}

  \begin{center}

   {\Large {\textbf{ Separated flow behind a backward-facing step.
  Part II. Experimental and numerical investigation of a turbulent flow.
    }}}

\vspace{0.5cm}

\textbf{Tatiana G. Elizarova$^1$, Eugene V. Shilnikov$^1$, \\
 R\'egine Weber$^2$, Jacques Hureau$^2$}

\vspace{0.5cm}

{$^1$ Inst. for Math. Modelling, Russian Academy of Sciences,\\
Miusskaya Square, 4a, Moscow 125047, Russia, elizar@imamod.ru \\
$^2$ Lab. de M\'ecanique et d'Energ\'etique, Polytech'Orl\'ans,\\
8, Rue L\'eonard de Vinci, 45072 Orl\'eans Cedex 2, France, Regine.Weber@univ-orleans.fr, Jacques.Hureau@univ-orleans.fr
}
\end{center}

\vspace{1cm}

We present the results of an experimental and numerical
investigation of a turbulent  flow over a backward-facing step in
a channel. Experimental data are visualized using a Particle Image
Velocimetry (PIV) device. As a mathematical model we used
quasi-hydrodynamic (QHD) equations. We have carried out numerical
modeling of the flow for three configurations that were also
studied experimentally. The computed flow proved to be
non-stationary. The averaged flow in the separation zone is in a
good agreement with the experimental data.

  \section{Introduction}

In this paper we compare the results of the experimental and
numerical investigation of a turbulent flow with medium Reynolds
numbers over a backward-facing step in a channel.

An adequate description of a non-stationary flow in the separation
zone behind the backward step is a considerable problem both in
experimental and numerical aspects. As far as the numerical
methods are concerned, a great variety of approaches are
currently proposed to describe a turbulent flow.
From the experimental point of view, visualization is a difficult task due
to the great differences in the velocity of the flow in the main
stream and in the separation zone, which hampers simultaneous
visualization of the flow above the step and in the separation zone.
Additional dificulties arise from the nonstationary  character of
the flow.

 At present there is a significant number of
computations of flows over a backward-facing step in a channel
using the averaged Navier-Stokes equations (see \cite{Armaly} -
\cite{So} and literature cited there). The length of the
separation zone related to the height of the step $L_s/h$ in
turbulent regime practically does not depend on the Reynolds number
and varies - according to  estimations \cite{Armaly} - \cite{So} -
from 5 to 8 depending on the geometry of the problem.

Averaged equations aimed at obtaining stationary numerical
solutions, but the ambiguity in choosing the parameters of the model
may lead to non-stationary flows behind a step as well
\cite{Lasher92}. Meanwhile even in the case of a developed
turbulent flow the question of validity of the turbulence
model remains. Non-stationary turbulent flow regimes may be
obtained using the LES (Large-Eddy Simulation) or DNS (Direct
Numerical Simulation) approaches, e.g. \cite{Fureby99} -
\cite{Meri}. At present there is a number of results in numerical
3D-simulation of flows over a backward-facing step (e.g.
\cite{Armaly2002}).

The experiments described in this paper were carried out at LME in
a wind-tunnel; attention was mainly paid to the flow in the symmetry
 plane. The experimental device was adapted
for low velocity flows with relatively small Reynolds numbers,
that, nevertheless, provided turbulent regimes behind a
backward-facing step. The results were obtained with a Particle
Image Velocimetry (PIV) device.

The numerical simulation of the flow was done on the basis of
quasi-hydrodynamic (QHD) equations, proposed by Yu.V. Sheretov
\cite{Sher2000}. QHD equations differ from the Navier-Stokes
system in additional dissipative terms of a second order
in space.

\begin{figure}[htb!]
\includegraphics[width=.95\textwidth]{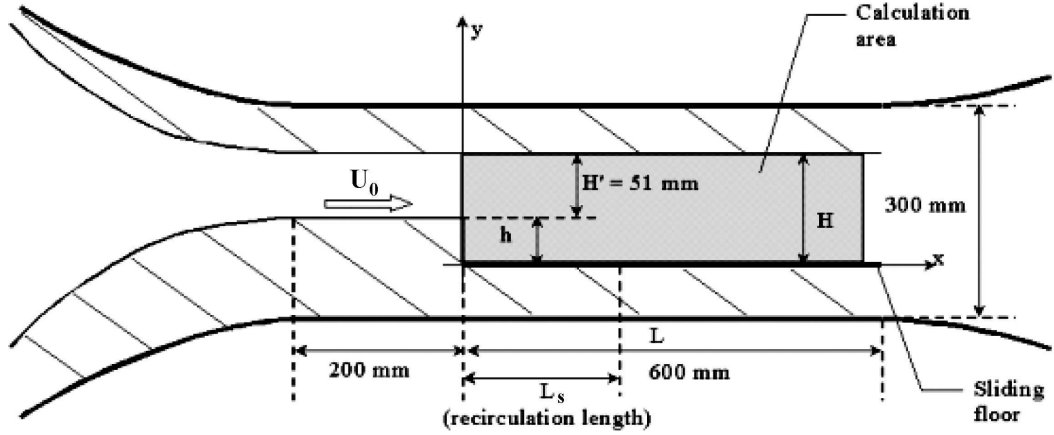}
\caption{Wind tunnel section and notations.}
    \label{Notations}
\end{figure}

The scheme of the investigated flow is presented in
Fig.\,\ref{Notations}. The Reynolds number is given by $Re
= (U_0h)/\nu$, where $U_0$ -- is the average gas velocity in the
entrance section of the channel, $h$ -- is the height of the step,
$\nu$ -- is the kinematic viscosity of the medium. The first
computational and experimental results were presented in
\cite{Rap2001}-\cite{ToulouseBail2004}. A brief description of
the QHD  system and its testing by computing a laminar
backward-facing  flow  is given in the part I of the present
publication \cite{Part1}.

\section{Experimental setup}

\subsection{Wind tunnel}
The experimental results have been obtained in an open air-driven
wind-tunnel at LME (Orl\'eans, France). The existing square section
($300\times 300$\,mm) has been adapted to present a
two-dimensional backward-facing step with adjustable step height.
The width and the length (800 mm) of the test section were not
modified. The electrically powered fan allowed us to obtain
velocities of about 1 to 80 m/s in the non-adapted test section
without any obstacle. Previous works published in the literature
about this subject \cite{Armaly} pointed out that the reattachment
length $L$ of the flow behind the step varies with the $Re$ number
and the height $h$ of the step: $L$ can be twenty times larger
than $h$. So, to be able to visualize the whole flowfield (and
especially the reattachment area), the test section height $H'$
upstream of the step was fixed to 51\,mm and we have considered
different step heights between 12\,mm and 50\,mm. The length of
the test section before the step is 200\,mm, which allows the flow to
be fully developed. The mean velocity obtained is approximately
1.4 m/s (as checked with a two-components hot-wire
anemometer). The mean turbulent ratio then measured is less than
0.85\,\% ahead of the step. One wall is transparent, for direct
visualization, and the other ones are black. A laser sheet enters
the wind tunnel section through a glass window in the upper wall.

\subsection{Particle Image Velocimetry (PIV)}
The experimental data generated are the two-dimensional components
of the velocity behind the backward-facing step. An
oil generator, placed at the entrance of the tranquillization room
(velocity about 0.01\,m/s), is used to generate and supply tracer
particles. The mean diameter of the particles is about 1\,$\mu$m.
The laser sheet is generated by a double-oscillator laser: a
Nd/Yag laser (Spectra Physics 400) adjusted on the second harmonic
and emitting two pulses of 200~mJ each ($\lambda=532$\,nm), at a
repetition rate of 10\,Hz.  The laser sheet is developed with an
optical arm containing mirrors. Lenses allow us to obtain a
laser sheet with a divergence of about 60$^\circ$ and with a
thickness of about 1~mm in the vicinity of the step.  For the present
experimental data, the flow images are picked up by a PIVCAM CCD
camera with $1008\times 1016$ sensor elements, placed
perpendicularly to the laser sheet. The laser pulses are
synchronized with the image acquisition by a TSI synchronizer
system driven by the InSight-NT$^{TM}$ software. As stated
previously, the reattachment length can be as long as twenty times
the height of the step. The aim of these experimental measurements
is to obtain the mean flow behind the step, so we have chosen to
decompose the flow area into several visualized areas ($80\times
80$\,mm or $135\times135$\,mm) to ensure a good precision in the
wake of the step. This is only possible because the mean velocity
field is studied. These sub-areas overlap by about 10\,mm. For all
the results presented here, the PIV recordings are divided into
interrogation areas corresponding to $64\times64$ pixels.  For
data post-processing, the interrogation areas overlap by 50\%.
The local displacement vector is determined for each interrogation
area by statistical methods (auto-correlation).  The projection of
the local flow velocity vector onto the laser sheet plane is
calculated by InSight using the time delay between the two
illuminations ($\Delta t$~=~1~ms) and the magnification at
imaging.  The post-processing used here is very simple -- no more
than a velocity range filter.

\section{Numerical modeling of turbulent flows}

Numerical modeling of flow over a backward-facing step  is carried
out in accordance with the experimental parameters for the
Reynolds numbers $Re(h)=4667$, $4012$, $1667$ and the ratios $h/H
= 0.5$, $0.44$, $0.33$ respectively.

The flow in the wind-tunnel is air

 at room temperature, atmospheric pressure with entrance
velocity $U_0$ $\sim$ 1.2 - 1.4 m/sec. The  sound velocity in
air under normal conditions equals $c_s$= 340 m/sec. Because
$U_0/c_s \sim 0.003$, we apply the approximation of a
viscous incompressible isothermal flow \cite{Lan}. Experimental
visualization is done in the symmetry plane; therefore, we use
a plane two-dimensional model of a flow in the numerical
computations.

These simplifying assumptions permit us to apply the finite
difference algorithm described and tested in
\cite{Part1} and allowing for non-stationary flows.
Unlike \cite{Part1}, we now specify a plane
velocity profile in the entrance section, that coincides with the
conditions of the experiment.

Numerical calculations were performed using the quasi-hydrodynamic
(QHD) equations that differ from
 the Navier-Stokes  ones by additional dissipative terms. These terms include a multiplying
 factor $\tau$, - relaxation (smoothing) parameter,
which has a dimension of  time.
 For laminar flows, $\tau$ is related
to the molecular viscosity coefficient.
 For turbulent flows the value of $\tau$ is no longer related with molecular viscosity and must be adjusted to fit
  the general flow features.
 The theory  of  QHD equations is presented  in \cite{{Sher2000}} and briefly described in \cite{Part1}.

The results of computations are gathered in Tables \ref{tabl1}--\ref{tabl3}
and are presented in Figs. \ref{Fig2}--\ref{Fig16}.
Additional computations may be found in \cite{MGU2003}. Tables
contain the serial number of computation (run), the dimensionless
value of the smoothing parameter $\tau$, the spatial
grid step, the number of grid nodes $N_y \times N_x$,
the dimensionless computing time $T_0$ and the reference to the figures
that illustrate this computation. Spatial grids are uniform and
have equal steps in both directions ($h_x = h_y$). Parameter $\tau$ varies
widely. The dimensionless time step in all presented variants is
equal to $\delta t= 10^{-4}$.

Velocity and pressure fields were recorded every 5000 steps, that
is, with a time interval $\Delta t_1=0.5$. The velocity components
in four particular locations behind the step were recorded every 500 steps,
i.e., with a time interval $\Delta t=500\,\delta t=0.05$.
Energy spectra of the velocity components pulsations were computed
for $Re=4667$ and $Re=4012$ according to the algorithm described in
the Appendix.

Computations were carried out in the dimensionless variables
 defined in \cite{Part1}. For $Re=1667$, we had $H=71$\,mm,
$U_0=1.25$\,m/s, and the dimensionless time $\tilde t =1$ corresponding
to $H/U_0=0.056$\,s; for $Re=4012$ and 4667, this variable corresponded
0.063\,s and 0.072\,s, respectively.

\subsection{Variant $Re=4667$}

\begin{figure}[htb!]
\begin{center}
\includegraphics[width=.65\textwidth]{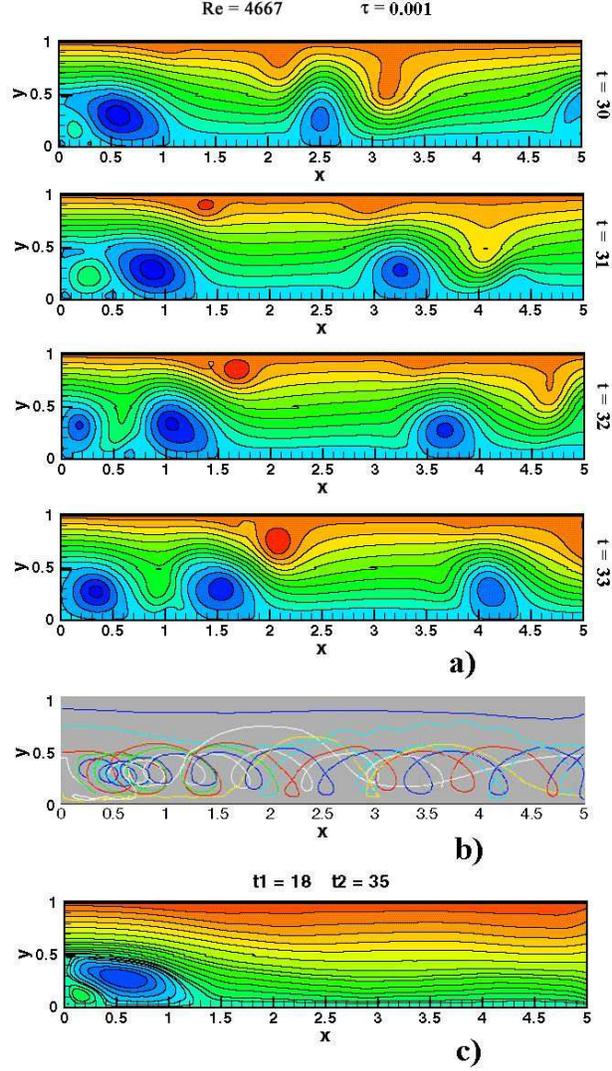}
\caption{Instant stream functions (a), trajectories of fluid
particles (b), averaged flow field (c) for Re=4667, run 2.}
    \label{Fig2}
\end{center}
\end{figure}

\begin{figure}[hbt!]
\begin{center}
\includegraphics[width=.65\textwidth]{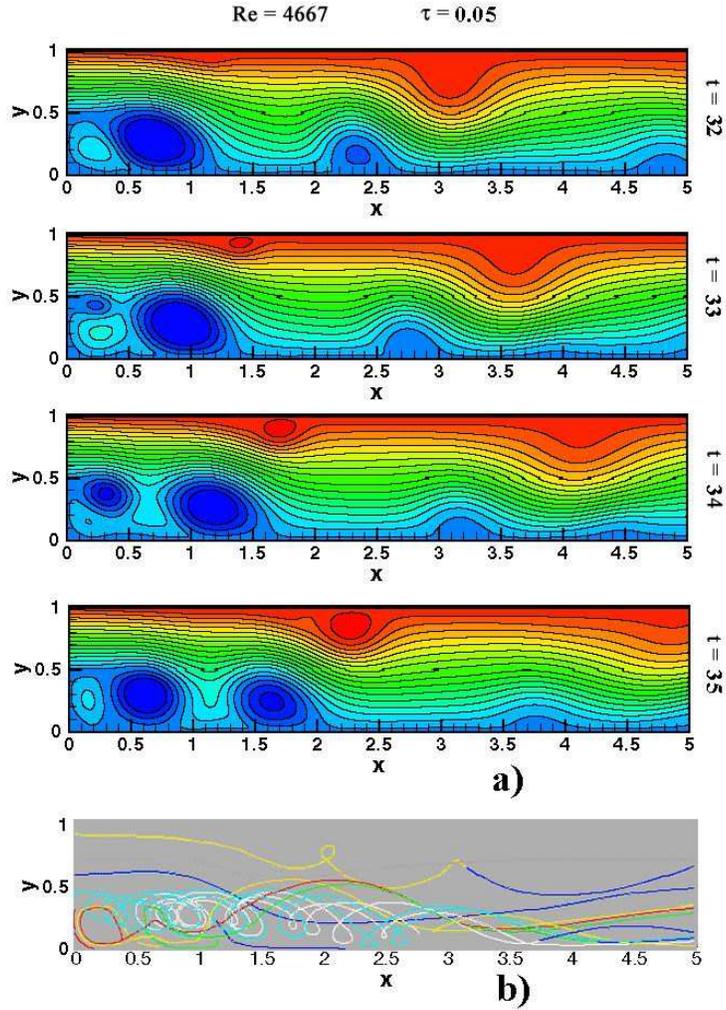}
\caption{Instant stream functions (a) and trajectories of fluid
particles (b) for $Re=4667$, run 4.}
    \label{Fig3}
\end{center}
\end{figure}

The major computational work for the proper choice of the
smoothing parameter  $\tau$ was carried out for $Re = 4667 $
(Tab.\ref{tabl1},   Figs.\ref{Fig2}--\ref{Fig8}).

\begin{table}[hbt!]
\medskip
\begin{center}
\begin{tabular}{ccccccc}
\hline N run& $\tau$  & $h_x=h_y$ &$N_y \times N_x$ & $L$ & $T_0$& Figure number
\\
\hline
1 & 0.0001 & 0.0125 & 80 x 400 & 5 & 20 &  -- \\
\hline
 2 & 0.001 & 0.00833 & 120 x 600 & 5 & 40 &  Fig.\ref{Fig2}  \\
 \hline
3 & 0.001 & 0.0125 & 80 x 400 & 5 & 20 &  --  \\
\hline
4 & 0.05 & 0.00833 & 120 x 600 & 5 & 120 &  Figs.\ref{Fig3} - \ref{Fig8}  \\
\hline
5 & 0.05 & 0.0125 & 80 x 400 & 5 & 120 &     \\
\hline
 6 & 0.1 & 0.0125 & 80 x 600 & 7,5 & 40 & -- \\
  \hline
    \hline
\end{tabular}
\end{center}
\caption{Computations for $Re=4667$, $H=101$\,mm, $U_0=1.40$\,m/s, $h/H=0.5$.}
\label{tabl1}
\end{table}
Fig. \ref{Fig2}a demonstrates the sequences of the stream function
isolines, computed according to algorithm \cite{Part1}, at
four successive times (indicated at the
right of the figure. A single
vortex appears behind the step; its length increases and then the
 single large vortex splits into several small ones that rotate in
 opposite directions. These smaller vortices are torn away from
the main one and carried down the flow till they leave the
computational domain.

Fig. \ref{Fig2}b demonstrates the trajectories of the fluid
particles; the Fig. \ref{Fig2}c demonstrates the velocity field time-averaged between $t_1$ and $t_2$ (values indicated in the figure).
The averaged velocities $u_x^{av}$ and $u_y^{av}$ in every
computational point are computed as
$$
    u_x^{av}=\frac{1}{t_2 -
  t_1}\int_{t_1}^{t_2}u_x(t)dt, \quad u_y^{av}=\frac{1}{t_2 -
  t_1}\int_{t_1}^{t_2}u_y(t)dt.
$$

The stream function is then computed using the averaged values. This procedure
 is correct due to the commutativity of
temporal averaging and spatial differentiation.

From Fig. \ref{Fig2} it is easily seen that for a quickly changing
flow the flow pattern looks essentially different at four different times.
 The recirculation flow is well
seen only on the stream functions that are constructed using
the averaged velocities (Fig.\,\ref{Fig2}c). The size of the
separation zone equals $L_s/h \sim 3$, which is twice less than
observed experimentally. The same mean size of the separation zone
is obtained in computations on a twice less accurate spatial
grid with the same $\tau=0.001$ (run 3) and also with asmaller $\tau = 0.0001$
(run 1). In the last case the time step was equal to $\delta t =
10^{-5}$. Therefore, for $\tau \le 0.001$ a quasi-periodical flow
is formed behind the step, and its shape rather weakly depends on
the value of $\tau$ and on the size of the spatial grid. In these
computations the separation zone proves to be smaller than the one
observed in experiments.

Computations 4 and 5 correspond to $\tau = 0.05$. In Fig.
\ref{Fig3}--\ref{Fig8} we demonstrate the results of computation 4.
As well as in other computations, instantaneous pictures of
the stream function (Fig. \ref{Fig3}a) look much the same and
strongly resemble those in Fig. \ref{Fig2}a. The trajectories of
fluid particles (Fig. \ref{Fig3}b) do not exhibit a recirculation zone.

\begin{figure}[htb!]
\begin{center}
\includegraphics[width=.7\textwidth]{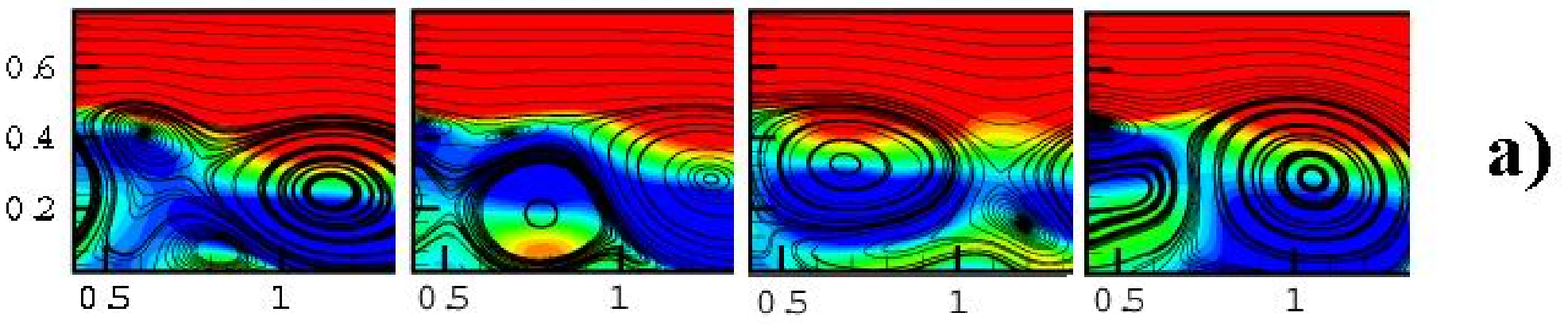}
\end{center}
\end{figure}
\begin{figure}[htb!]
\begin{center}
\includegraphics[width=.7\textwidth]{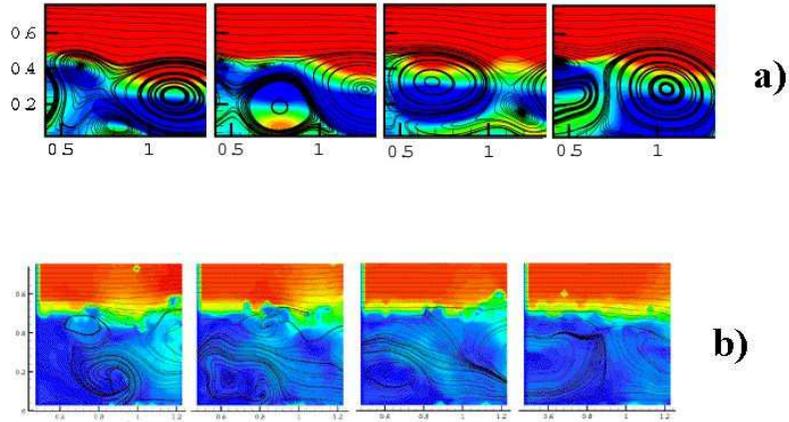}
\caption{Fragments of the averaged flow field: computational (a) and experimental (b).} \label{Fig4}
\end{center}
\end{figure}

In Fig.\ref{Fig4}a we show computed pictures of a flow fragment behind the step. This fragment corresponds
to the region 0 - 0.75 along the ordinate and 0.4 - 1.3 along the abscissa. Each numerical pattern  is
constructed from the instantaneous velocity vectors,
 averaged over time intervals  $t_2 - t_1=0.015$ (1 ms).
   We give 4 successive pictures
starting from times $T_1=70$, $T_2=71,5$, $T_3=73$ and
$T_4=74,5$.
 So the numerical flowfields are obtained at a time interval of 1.5 (0.1 s, frequency 10 Hz).

 Fig.\ref{Fig4}b shows the time sequence of experimental flowfields obtained at a frequency of 10 Hz in the same part
  of the flow domain behind the step. The velocity  fields are obtained by intercorrelation between
  two images acquired with an interval time of 1 ms, that corresponds with the time delay between two laser
  illuminations.

Averaging intervals in Figs.\ref{Fig4}a and \ref{Fig4}b are approximately the same. Experimental and
computational portraits of the flow both demonstrate a chaotic non-stationary kind of flow, and the typical
sizes of inhomogeneities obtained from computations agrees with the ones observed experimentally.

\begin{figure}[htb!]
\begin{center}
\includegraphics[width=.75\textwidth]{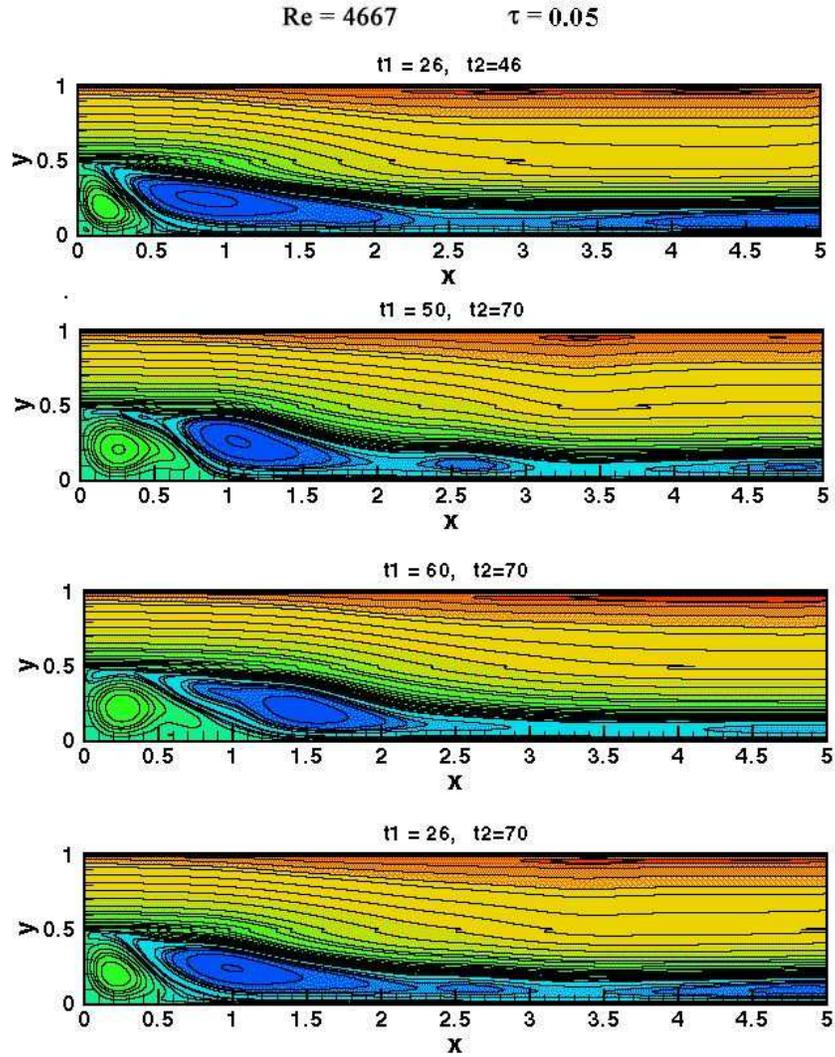}
\caption{Averaged flow fields, computation, $Re=4667$.}
    \label{Fig5}
\end{center}
\end{figure}

Fig. \ref{Fig5} we show the computed
flow fields, averaged over larger time intervals ($t_2 - t_1$
equals 0.5 - 2 s). In Fig. \ref{Fig6} we show
the experimental flow fields averaged over
intervals of 8-10 seconds. One can see the good agreement - both
in size ($L_s/h \sim 6$) and in structure of the separation zone
in these averaged pictures. In both cases the averaged pictures
depend on initial time $t_1$ and integration time $t_2-t_1$, but this dependence is weak, because the averaging time is essentially larger than the typical time of velocity fluctuations.

\begin{figure}[htb!]
\begin{center}
\includegraphics[width=.75\textwidth]{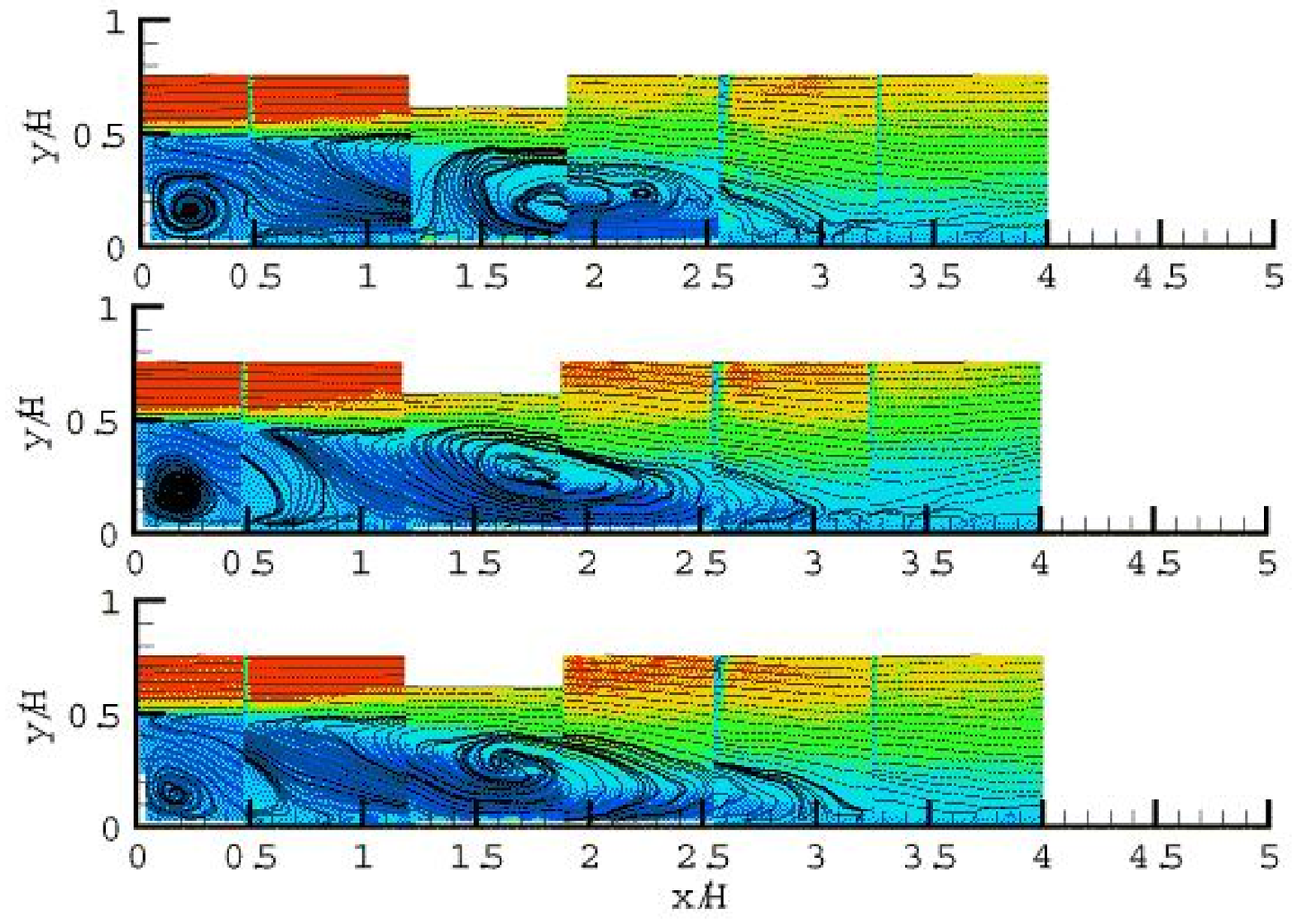}
\caption{Averaged flow fields, experiment, $Re=4667$.}
    \label{Fig6}
\end{center}
\end{figure}

Fig. \ref{Fig7} illustrates the time dependence of velocity
components $u_x$ and $u_y$ as functions of time at three points
located behind the step and far from the wall. The flow is
essentially non-stationary and quasi-periodical. The general
features of these curves depends weakly on the location of the
points.

\begin{figure}[htb!]
\begin{minipage}{.48\textwidth}
\includegraphics[width=\textwidth]{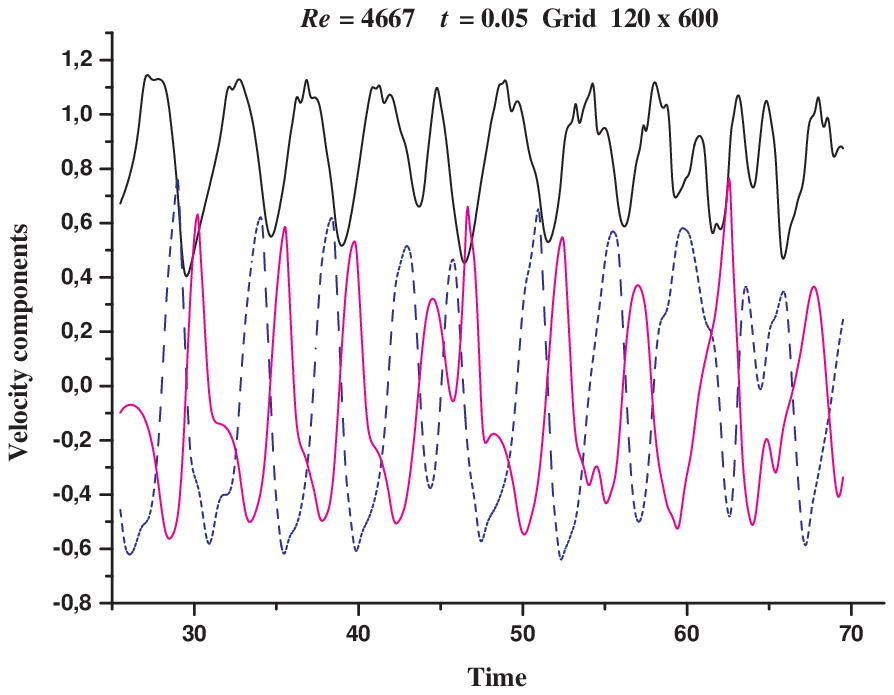}
\caption{Fragment of temporal evolution of the velocities,
computation, $Re=4667$.}
    \label{Fig7}
\end{minipage} \hfill
\begin{minipage}{.48\textwidth}
\includegraphics[width=\textwidth]{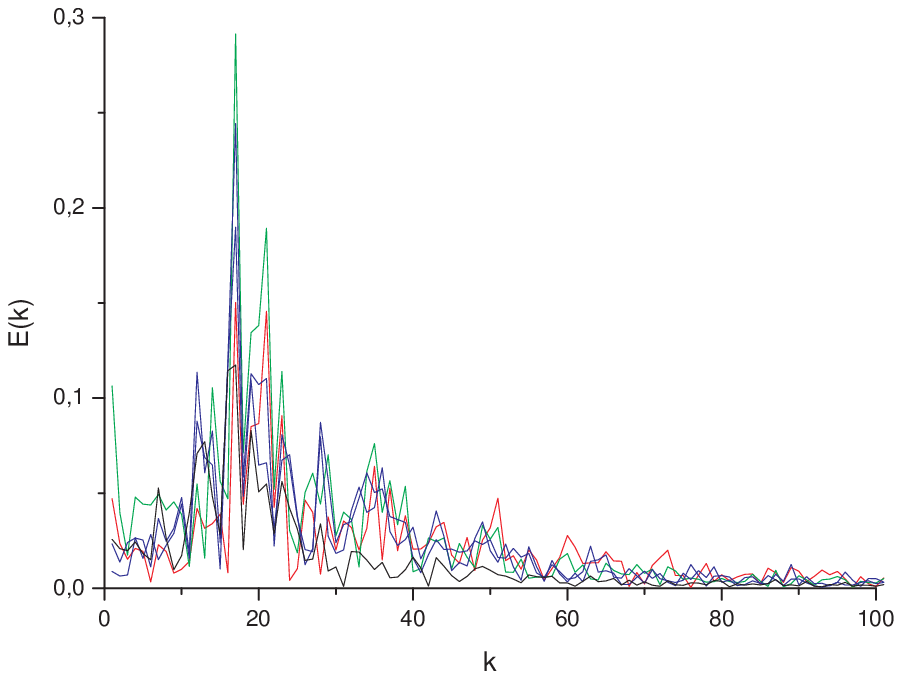}
\caption{Velocity pulsation spectrum $E(k)$, computation, $Re=4667$.}
    \label{Fig8} \end{minipage}
\end{figure}

The pulsation spectra $E(k)$ of the velocity component $u_x$ at three
 spatial points are plotted in Fig. \ref{Fig8}. The mentioned
spectra may be considered as reliable up to $k\approx 240$. The $k=0$
harmonic is absent because it is related to the average velocity
$\bar u$. Fourier series expansion is carried out after completing
the initial relaxation to stable oscillations during the time
interval from $t = 40$ to $t=120$. The main oscillation
frequencies and the decrease of amplitude for the harmonics at large
and small values of $k$ are easily seen.

The results of computations 4 and 5 on different spatial grids
were shown to be close to one another.

Computation 6 is carried out with a large value of the smoothing
 parameter $\tau $.
The structure of the flow proves to be non-physical - the
non-stationary separation zone behind the step grows indefinitely
and reaches the right border of the computation zone. In this case
the velocities in the flow change slowly and the trajectories of
the fluid particles are close to the streamlines.
The averaged flow is of no interest in this computation.
Additional computations and pictures illustrating them can be
found in \cite{MGU2003}.

\subsection{Variant $Re=4012$}

Computation results for this variant in general repeat those
obtained before. In run 1 (see Table 2)
we obtain a non-stationary solution with a separation zone
length ($L_s/h \sim 2$) smaller than the experimental one.

  \begin{table}[ht!]
  \medskip
  \begin{center}
  \begin{tabular}{ccccccc}
  \hline N run& $\tau$  & $h_x=h_y$ &$N_y \times N_x$ & $L$ & $T_0$&
   Figure numbers
  \\
  \hline
   1 & 0.001 & 0.0125 & 80 x 400 & 5 & 20 & --   \\
  \hline
   2 & 0.05 & 0.00833 & 120 x 600 & 5 & 200 & Figs.\ref{Fig9b} - \ref{Fig14}  \\
   \hline
    \hline
  \end{tabular}
  \end{center}\caption{Computations for $Re=4012$, $H=92$\,mm, $U_0=1.45$\,m/s, $h/H=0.44$.}
  \label{tabl2}
  \end{table}

\begin{figure}[htb!]
\begin{center}
\includegraphics[width=.8\textwidth]{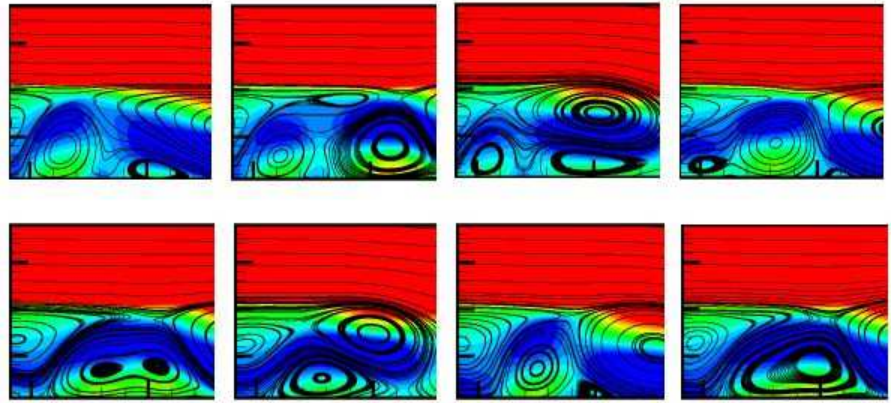}
    \label{Fig9a}

\centerline\huge{a)} \vspace{5mm}

\includegraphics[width=.8\textwidth]{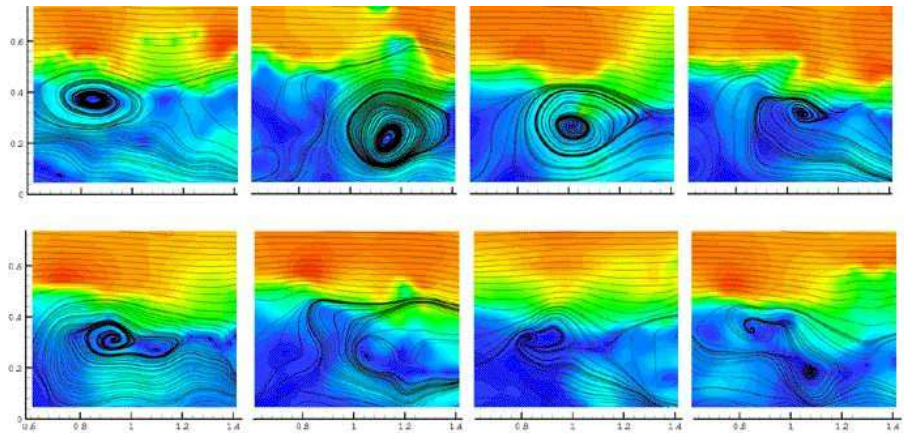}

\centerline\huge{b)} \vspace{5mm}

\caption{Fragments of the averaged flow field: computational (a)
and experimental (b) for $Re= 4012$.}
    \label{Fig9b}
\end{center}
\end{figure}

In computation 2 with $\tau =0.05$ we get pictures of the flow
averaged over $t_2 - t_1$ = 0.015  (see Fig. \ref{Fig9b}a). In this figure, we
show 8 successive pictures starting from T=154 with a time
interval 1.5, that corresponds to a frequency of 10Hz. The
corresponding experimental flow fragments are given in Fig.
\ref{Fig9b}b starting from time $t_1=0.4$\,s with an
acquisition rate equal also to 10Hz. Both starting points are chosen
arbitrarily. Both calculated and experimental flow fields exhibit a chaotic non-stationary character.

\begin{figure}[htb!]
\begin{center}
\includegraphics[width=.7\textwidth]{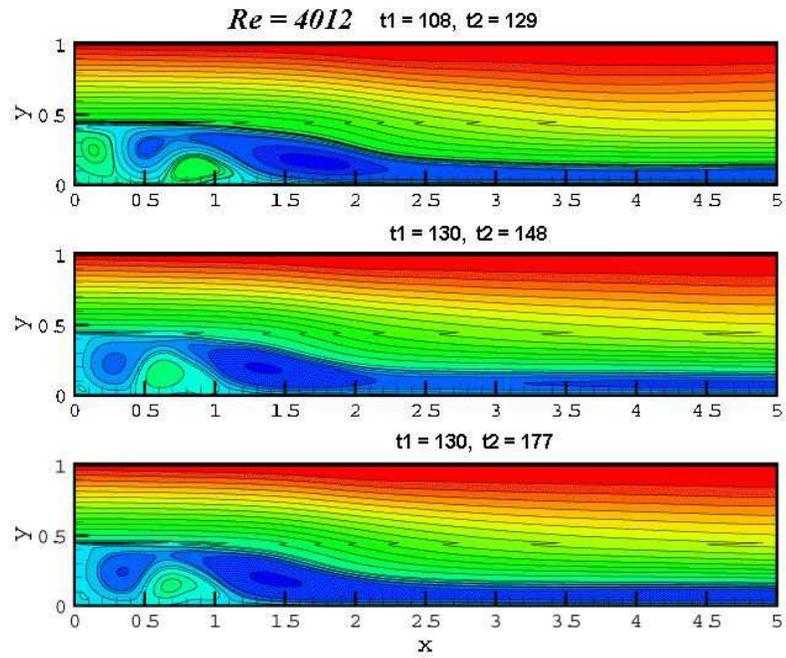}
\caption{Averaged flow fields, computation, $Re=4012$.}
    \label{Fig10}
\end{center}
\end{figure}

\begin{figure}[htb!]
\begin{center}
\includegraphics[width=.8\textwidth]{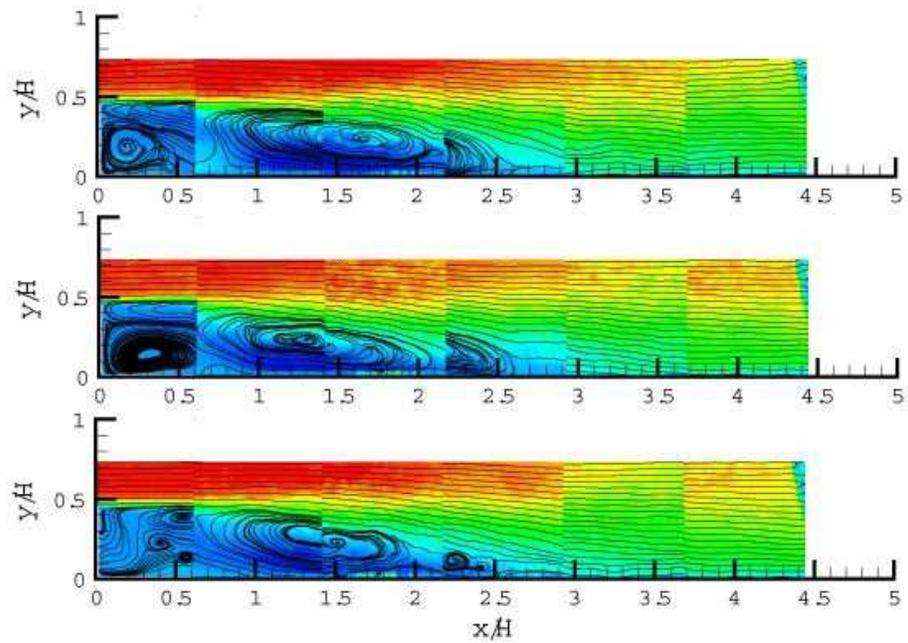}
\caption{Averaged flow fields, experiment, $Re=4012$.}
    \label{Fig11}
\end{center}
\end{figure}

In Fig. \ref{Fig10} we demonstrate the streamlines,
averaged over long time intervals and constructed in the same way as in
\cite{Part1}. In Fig. \ref{Fig11} the corresponding experimental
pictures of the flow in the separation zone are shown over a
background color that represents the velocity component $u_x$.


\begin{figure}[htb!]
\includegraphics[width=.8\textwidth]{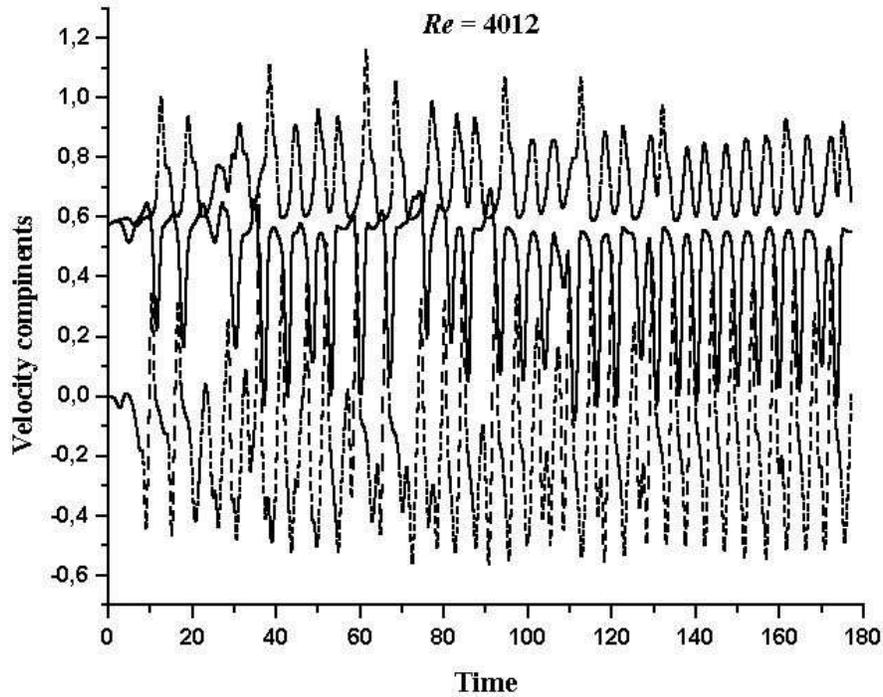}
\caption{Temporal evolution of the velocity behind the step,
computation, $Re=4012$.}
    \label{Fig12}
\end{figure}

\begin{figure}[htb!]
\begin{minipage}{.49\textwidth}
\includegraphics[width=\textwidth]{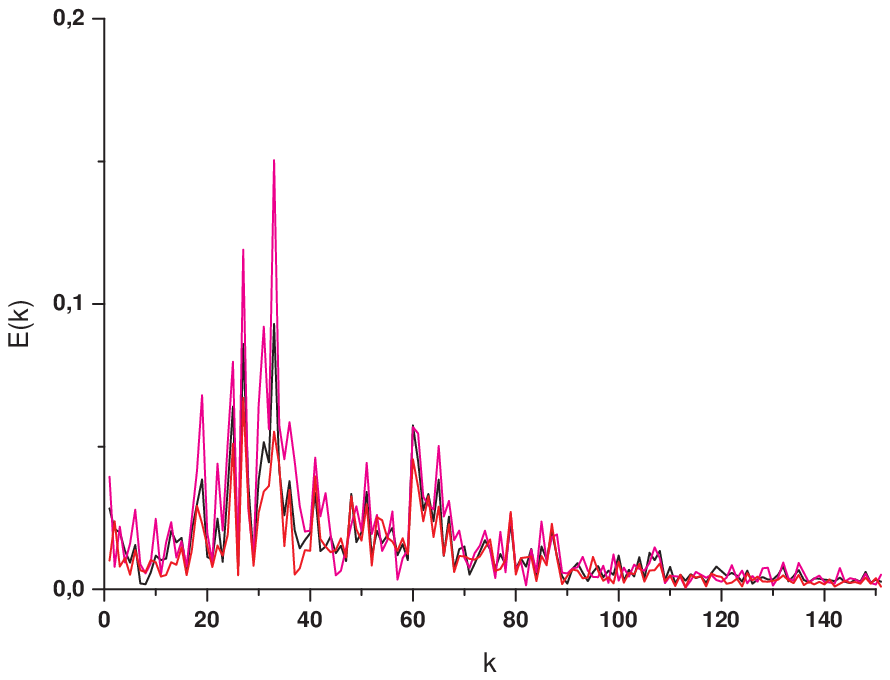}
\caption{Velocity pulsation spectrum E(k), computation, $Re=4012$.}
    \label{Fig13}
\end{minipage} \hfill
\begin{minipage}{.49\textwidth}
\includegraphics[width=\textwidth]{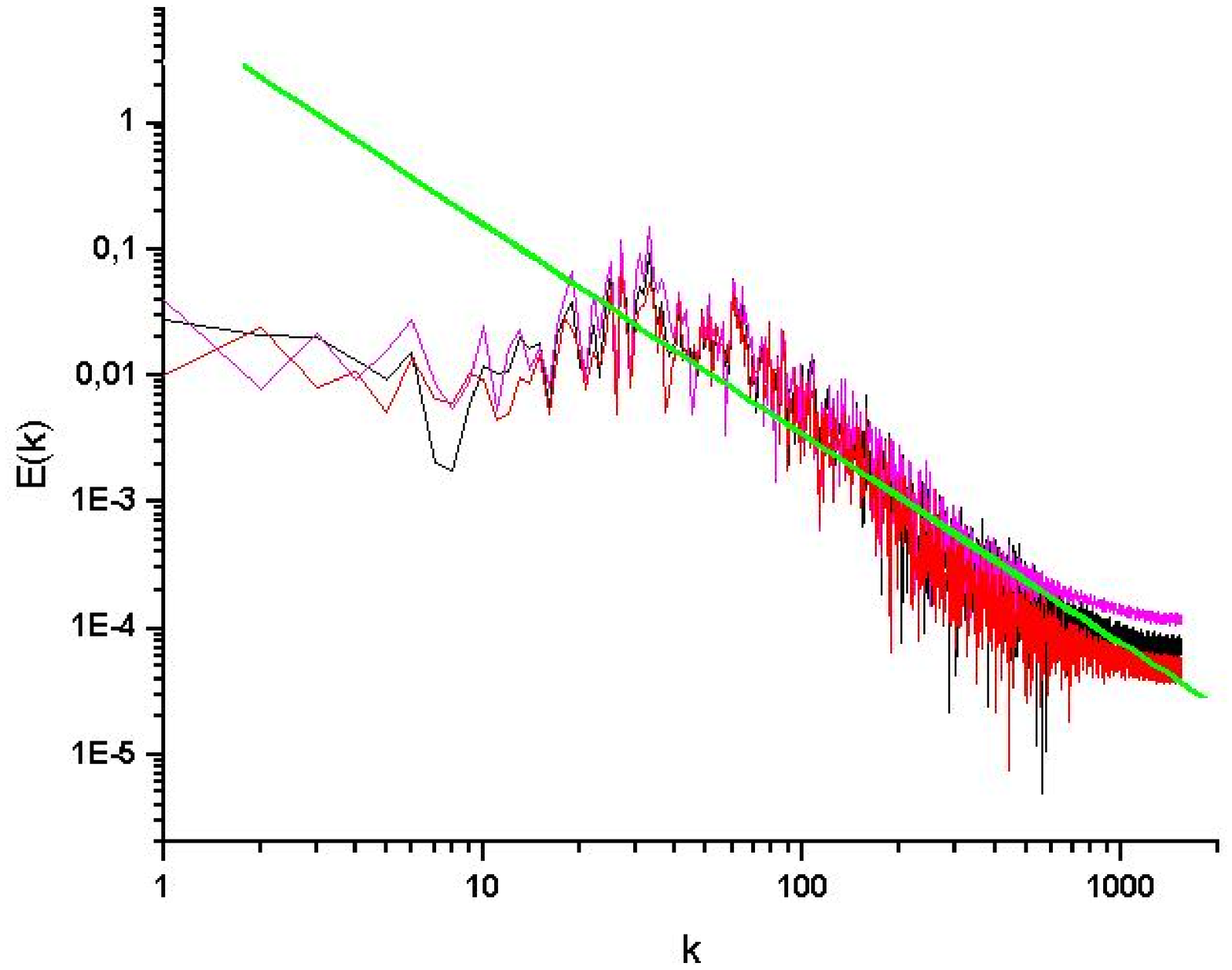}
\caption{Velocity pulsation spectrum E(k), Logarithmic scale,
computation, $Re=4012$.}
    \label{Fig14}
\end{minipage}
\end{figure}

Fig. \ref{Fig12} depicts the time evolution of the velocity
components behind the step. It demonstrates quasi-periodical
oscillations. Figs. \ref{Fig13} and \ref{Fig14} show the energy
spectrum of pulsations of the velocity components in linear and
logarithmic scales, respectively. The last one contains the
dependence $E(k) \sim k^{-5/3}$ (Kolmogorov-Obukhov law in
spectral form) \cite{Lan}. This law of the kinetic energy
dissipation is typical for well-developed turbulent flows. It can
be easily seen that the numerical energy spectrum in the frequency
range, that agrees with the step of the spatial grid and with the
time of computation $T_0$, coincides in general with the classical
law of the decreasing of energy $E$ with the growth of the wave
number $k$.


  \subsection{Variant $Re=1667$}

For $Re = 1667$ (see Table 3), in run 2, the length and the
structure of the separation zone ($L_s/h \sim 6$) agrees with the
experimental data. In run 1 ($L_s/h \sim 2,5$) the length
of the separation zone proves to be smaller than in the
experiment. In run 3 the numerical solution becomes
non-physical - the separation zone grows indefinitely.

  \begin{table}[ht!]
  \medskip
  \begin{center}
  \begin{tabular}{ccccccc}
  \hline N run& $\tau$  & $h_x$ &$N_y \times N_x$ & $L$ & $T_0$&
   Figure number
  \\
  \hline
   1 & 0.001 & 0.0125 & 80 x 400 & 5 & 20 & --   \\
  \hline
   2 & 0.02 & 0.0125 & 80 x 480 & 6 & 160 & Fig.\ref{Fig15}, \ref{Fig16}  \\
  \hline
   3 & 0.05 & 0.0125 & 80 x 480 & 6 & 60 &  -- \\
   \hline
    \hline
  \end{tabular}
  \end{center}\caption{Computations for $Re=1667$, $H=71$\,mm, $U_0=1.25$\,m/s, $h/H=0.33$.}
  \label{tabl3}
  \end{table}

\begin{figure}[htb!]
\begin{center}
\includegraphics[width=.6\textwidth]{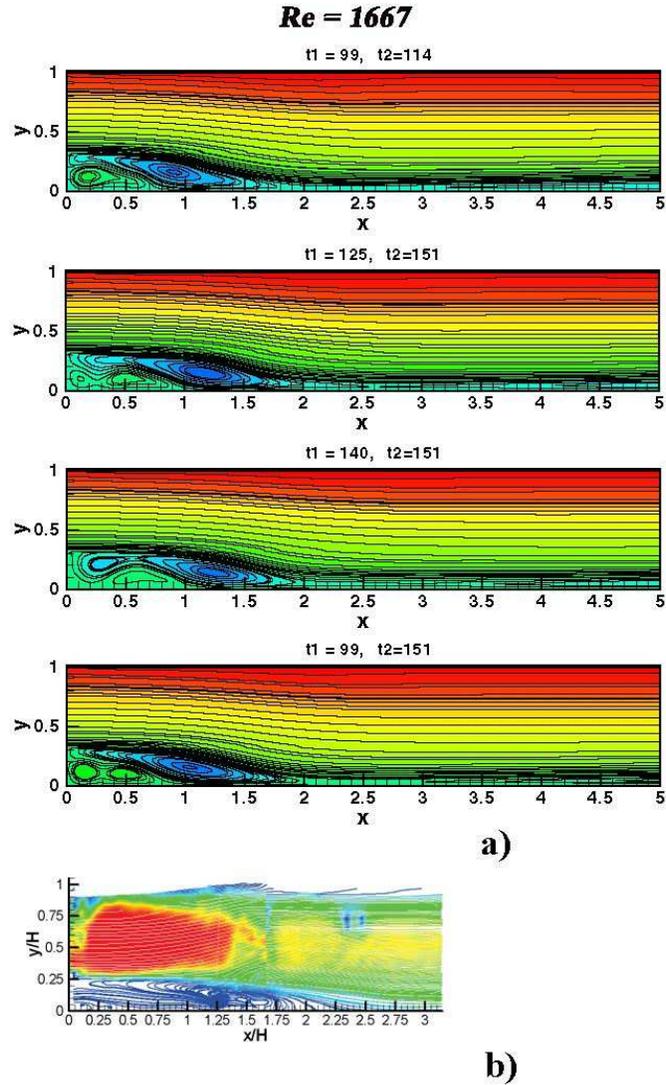}
\caption{Averaged flow fields: computational (a) and experimental
(b) for $Re=1667$.}
    \label{Fig15}
\end{center}
\end{figure}

\begin{figure}[htb!]
\begin{center}
\includegraphics[width=.65\textwidth]{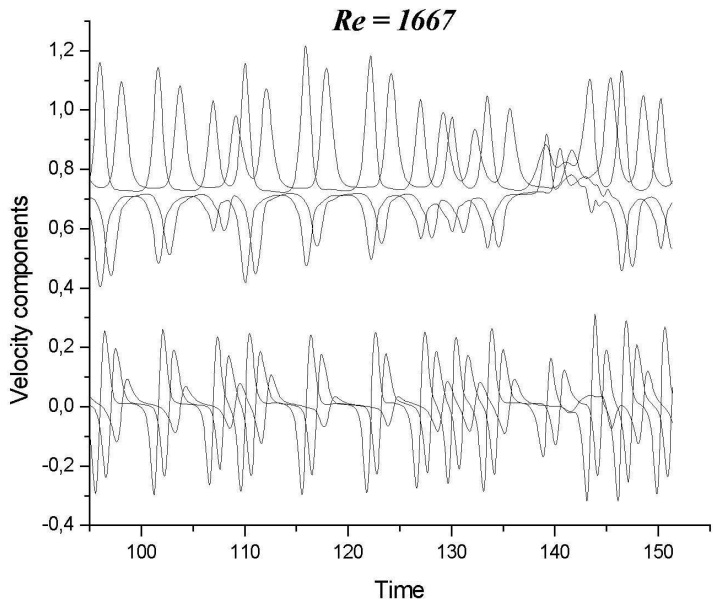}
\caption{Temporal velocity evolution behind the step, computation,
$Re=1667$.}
    \label{Fig16}
\end{center}
\end{figure}

In Fig. \ref{Fig15}a we show a series of the averaged stream
function fields (run 2) The corresponding averaged
experimental flow picture is shown in Fig. \ref{Fig15}b. The size and the
structure of the flow in the separation zone are in good
agreement. Fig. \ref{Fig16} illustrates the temporal evolution of
flow velocity components; it presents a quasi-periodical character.


  \section{Discussion and conclusions}

\subsection{Relaxation parameter}

Numerical computations are based on the QHD equations. They differ
from the Navier-Stokes system by additional dissipative terms that
contain a small parameter $\tau$ - relaxation (smoothing)
parameter. While computing turbulent flows, $\tau$  is chosen much larger
than for computing laminar flows
\cite{Part1}. We manage to obtain the structure of the flow behind a
backward-facing step by varying $\tau$. The dependence of
the averaged flow on $\tau$ proves to be relatively weak: in the
variant $Re=4667$ the  length of the separation
zone varied from $L_s/h \sim 3$ to $L_s/h \sim 6$, when $\tau$ was varied from 0.001 to 0.05. We should also mention that the parameter
of relaxation $\tau$ is the only constant in the QHD model that
must be adjusted to achieve agreement between numerical and
experimental results.

A posteriori specification of the relaxation parameter on the
basis of the flow field resembles to adjusting the turbulent
viscosity in phenomenological models of turbulence. In these
models the value of turbulent viscosity is not related with the
coefficient of the molecular viscosity. This quantity is derived
from experimental data, e.g., from the pulsation features
of the flow or from its characteristic scales \cite{Lan}.

Agreement with the experimental data - as far as the length and
the structure of the separation zone are concerned - is obtained
in variants $Re=4667$ and $Re=4012$ for $\tau = 0.05$, and in variant
$Re=1667$ for $\tau = 0.02$. In dimensional form it corresponds to $3.6 \cdot
10^{-3}$\,s, $3.2 \cdot 10^{-3}$\,s and $1.12 \cdot 10^{-3}$\,s
respectively. In these cases the average length of the separation
zone equals $L_s/h \sim6$, and the position of the vortex centers
coincide in computations and in experiments. The position
of  a secondary vortex located at the internal corner is also well
described in the computations.

Mention, that in our calculations we have not seen the well-known
anomaly of low-Reynolds-number (LRN) Reynolds-stress models (RSM)
(\cite{Shima}, \cite{So}), that is a spurious bending (backward
curvature) of the dividing streamline at the reattachment point.

There remains the problem of a priori choice of the relaxation
parameter in QHD equations for computing turbulent flows.
Nevertheless, we may claim that $\tau \le 1$, because the
dissipative terms should be considerably smaller than the
convective ones. In computations the minimal resolution scale is
the step of the spatial grid. Therefore we may suppose that $\tau$
should be greater than the characteristic perturbation spreading
time over one spatial step $h_x$; that is, $\tau\ge h_x/c_s$.
Besides, for the problem under consideration, $\tau$ should be a
fortiori greater than the averaging time, for which the velocity
profile at the entrance of the channel may be considered as a
flat. This time was not measured in the present experiments.

\subsection{Flow visualization and time averaging intervals}

The numerically observed type of the flow formation resembles the
well-known Landau turbulence scenario \cite{Lan}. This formation
consists of the growth of initial perturbations that are
suppressed by nonlinear media reactions. It leads to the
appearance of a complex quasi-periodical motion, which seems to be
turbulent[JCL6].

The computational  time step $\delta t$ that describes the
variation of the nonstationary velocity field, is several orders of
magnitude smaller than the time resolution of the experiments. So it is
impossible to compare  the numerical and experimental flow
patterns  directly, using instantaneous streamlines. The trajectories of the liquid particles, based on the stream-functions, are not suitable for the
comparison with  the present experiments. That is why we used
averaged velocity fields when comparing numerical and experimental
results.

We used two types of temporal intervals for visualizing the
results of computations - large intervals for comparing global
flow pictures and small intervals for comparing local "snapshots".

The averaging interval for global pictures equals $t_2 - t_1
= 10 - 50$ dimensionless units, that is, 0.5-3\,s. This time
interval exceeds largely the averaged period of the temporal
velocity evolution (Figs. \ref{Fig7}, \ref{Fig12} and
\ref{Fig16}). The number of averaged fields proves to be large
enough, because the flow recording is done with the interval  0.5.
In other words, the number of averaged velocity fields in
computation approximately equals the number of averaged
experimental images. It makes both averaging procedures equivalent
in general. The obtained fields depend weakly on the beginning and
length of the time averaging interval.

The averaging interval for comparing numerical and experimental
"snapshots" equals to 0.015 dimensionless units, approximately
1\,ms, which coincides with the time delay between the two
illuminations - a gap between two pulses of the PIV installation.
 The numerical flowfields are obtained at a time interval $\sim 0.1$\,s., to reproduce the experimental
 acquisition frequency of frequency $\sim 10$\,Hz.
The computational and the experimental averaged pictures in
general look alike - both consist of small chaotically mixed
non-stationary whirls that essentially depend on the choice of the
temporal interval $t_2 - t_1$.

It is impossible to observe instant velocity fields
experimentally. Therefore the question of validity of the computed
instant velocity fields is physically senseless. At the same time
numerical and experimental fields averaged over large and small time
intervals prove to be consistent with each other.

\subsection{Conclusions}

In \cite{Part1} the QHD equations were tested in a laminar flow
simulation. There a relaxation parameter $\tau$  was determined by
the molecular viscosity. In computations it was chosen as small as
possible to provide the stability of the numerical computations.
The stationary flow did not depend on its value. The initial
oscillations of the solution faded during the time.

In the present paper we demonstrate that in  the turbulent flow
the initial perturbations do not fade and lead to the formation of
non-stationary flow pattern, that resembles a periodical process.

This type of solution is observed for sufficiently small values of
the dimensionless temporal relaxation  parameter: $\tau \le 0.1$.

The comparison of numerical solutions with experimental data is
possible only on the basis of the averaged velocity fields. In
this case the outlook of the averaged flow depends both on the
value of the temporal relaxation parameter $\tau$ (which makes the
difference between the QHD system and the Navier-Stokes equations)
and on the chosen averaging interval.

In spite of considerable simplifications of the problem - using
a two-dimensional model of the viscous non-compressible
isothermal fluid - the results of computation are in good
agreement with the present experiments.

We suppose that the numerical modeling of the turbulent flow over
a backward-facing  step by means of the QHD system is possible due
to the additional (in comparison with the Navier-Stokes equations)
dissipative terms that are small for the stationary flows and
become not small for the non-stationary turbulent flows.

  \section {Appendix. Computing the spectrum of pulsations}

The energy spectrum of the flow pulsations is computed on the basis of
the time dependence of velocity components $u_x$ and $u_y$.
Let us write the velocity $u_x$ as $$u_x=\bar u_x + u_x',$$
where $\bar u_x$ is the average velocity value, $u_x'$ is its
pulsation component. Then we apply the discrete Fourier transform
to $u_x'$ (for simplicity further on we omit primes).
  We present the pulsation component in form of a series

  $$u_x=\sum_{k=1}^{N/2}(\frac{a_{k+1}}{N} cos
  \frac{2\pi lk}{N} + \frac{b_{k+1}}{N} sin \frac{2\pi lk}{N}).$$

  To find the spectrum of pulsations let us compute Fourier coefficients:

  $$ a_{k+1}=2\sum_{l=0}^{N-1} u_x^{l+1} cos \frac{2\pi
  lk}{N}, \quad b_{k+1}=2\sum_{l=0}^{N-1} u_x^{l+1} sin \frac{2\pi
  lk}{N}, $$ where $u_x^l$ - is the value of the velocity $u_x$ at
  $i,j$ at time $l\Delta t$; $N$ is the number of the considered records, $k$ runs from 1 to $N/2$.

  The energy spectrum is derived from these Fourier coefficients:

$$ E(k) = a_k^2 +b_k^2. $$
The maximal value of $k= N/2$ corresponds to oscillations of
period $\Delta t$. When the frequency of recordings is chosen,
oscillations with higher frequencies are not allowed. But for
proper resolution we need at least 4-5 points per period.
Therefore the obtained spectrum can be considered as reliable for
$k\le N/8$.

For all variants of the present calculations, $\Delta t =0.05$. For example, for $Re=4012$, run 2, $T_0 =
200$, so the reliable range of $k$ in Figs. \ref{Fig13}, \ref{Fig14} is limited to $k\le T_0/\Delta t /8 =
500$ approximately.

\section { Acknowledgement}
 Authors are thankful to  Jean-Claude
Lengrand (Lab. d'Aerothermique du CNRS, Orleans) for the fruitful
discussions and his contributions to the preparation of this
article.


\begin{thebibliography}{99}

   \bibitem{Armaly} Armaly B.F., Durst F., Pereira J.C.F., Schonung B.:
  Experimental and theoretical investigation of backward-facing
  step flow. J. of Fluid Mech.  V. 127. pp. 473--496, 1983.

  \bibitem{Nalla} Nallasamy M.: Turbulence models and their applications to the
  prediction of internal flows: a rewiew.  Computers and Fluids. V.
  15. N 2. pp. 151--194, 1987.


  \bibitem{Tong} Tong G.T.: Modelling turbulent recirculating flows in complex
  geometries. Computational Techniques and Applications: CTAC-83,
  Ed. Noye J., Fletcher C. Elsevier Science Publishers B.V.
  (North--Holland), pp. 653--668, 1984.


  \bibitem{Speziale} Speziale C.G., Ngo T.: Numerical solution of turbulent
  flow past a backward facing step using a nonlinear K-$\epsilon$
  model. Int. J. Eng. Sci., Vol.26, No 10, pp.1099--1112, 1988.

  \bibitem{Thangam} Thangam S., Speziale C.G.: Turbulent Flow Past a
  Backward-Facing Step: A Critical Evaluation of Two-Equation
  Models. AIAA Journal, Vol.30, No 5, pp.1314--1320, 1992.


  \bibitem{Lasher92} Lasher W.C., Taulbee D.B.: On the computation of
  turbulent backstep flow. Int. J. Heat and Fluid Flow, Vol 13, No 1,
   pp.30--40, 1992.

\bibitem{Shima} Shima N.: Low-Reynolds-number second moment closure
without wall-reflection redistribution terms. Int. J. Heat and
Fluid Flow. 1998, 1998. Vol.19, pp.549--555.

\bibitem{So} So R.M.C., Yuan S.P.: A geometry independent near-wall
Reynolds-stress closure. Int. J. Eng. Sci. 1999. Vol 37, pp.33--57.

  \bibitem{Fureby99} Fureby C.: Large Eddy Simulation of
  Rearward-Facing Step Flow. AIAA Journal, Vol.37, No 11, pp.1401--1410, 1999.

\bibitem{Le}Le H., Moin P., Kim J.: Direct numerical simulation of
turbulent flow over a backward-facing step. J. Fluid Mech.1997. Vol
330, pp.349--374.

  \bibitem{Meri}Meri A., Wengle H., Schiestel R.: DNS and LES of a
  backward-facing step flow using 2nd- and 4th-order spatial
  discretization and LES of the spatial development of mixing of
  turbulent streams with non-equilibrium inflow conditions. Notes on
  Numerical Fluid Mechanics.  Springer, V.75, 2000, pp. 268--287.

  \bibitem{Armaly2002} Armaly B.F., Li A., Nie J.H.:
  Three-Dimensional Forced Convection Flow Adjacent to
  Backward-Facing Step. Journal of Thermophysics and Heat Transfer,
  Vol.16, No 2, p.222--227, 2002.

   \bibitem{Sher2000} Sheretov Yu.V.: Quasi-hydrodynamic and quasi-gasdynamic
   equations based mathematical modeling of liquid and gas flows. Tver: Tver State University
  2000.

  \bibitem{Rap2001} Elizarova T.G, Kalachinskaya I.S., Weber R, Hureau J.,
   Lengrand J.-C.:   Ecoulement derri\`ere une marche. Etude exp\'erimentale et num\'erique.   Laboratoire d'A\'erothermique du CNRS,  Orl\'eans (Fr), R~2001 - 1 (2001).

  \bibitem{TOULOUSE2002}
   Elizarova T.G, Kalachinskaya I.S., Weber R, Hureau J., Lengrand J.-C.:
  Backward-facing step flow. Experimental and numerical approach.
  IUTAM Symposium "Unsteady separated flows". 8-12 April 2002, Toulouse,
  France. Abstracts.

\bibitem{MGU2003}Elizarova T.G., Kalachinskaia I.S., Sheretov Yu.V., Shilnikov E.V. (2003):
 Numerical simulation of the backward-facing step flows. Prikladnaia Matematika i Informatika:
 Trudy Faculteta VMiK MGU. Moscow, MAKS Press, 2003, no. 14, pp.
 85--118 (in Russian)

\bibitem{ToulouseBail2004}T.G. Elizarova, E.V. Shilnikov, R. Weber, J. Hureau, J.-C.
Lengrand.: Experimental and Numerical Investigation of the
Turbulent Flow behind a Backward-Facing Step. Int. Conference on
Boundary and Interior Layers (BAIL) 2004,  France, Toulouse,
5--7 July 2004. Proceedings on CD.

   \bibitem{Part1} Elizarova T.G., Kalachinskaya I.S., Sheretov
   Yu.V.:
 Separating flow behind a back-step. Part I.
Quasi-hydrodynamic equations and computation of a laminar flow.
http://arXiv.org/abs/math-ph/0407053 (2004).

  \bibitem{Lan} Landau L.D., Lifshitz E.M.: Hydrodynamics, Ed. Nauka, Moscow, 1986
 (in Russian)


  \end{thebibliography}
  \end{document}